\pdfoutput=1
\documentclass[preprint,preprintnumbers,amsmath,amssymb,floatfix,endfloats*]{revtex4}
\usepackage{graphicx}% Include figure files
\usepackage{dcolumn}% Align table columns on decimal point
\usepackage{bm}% bold math
\usepackage{amsfonts}

\DeclareMathAlphabet{\mathsfsl}{OT1}{cmr}{bx}{it}
\begin{document}
%----------------------------------------------------------------------%
% Title
%----------------------------------------------------------------------%
\title{Atomistic modeling of heat treatment processes for tuning the mechanical properties of disordered solids}
\author{Nikolai V. Priezjev$^{1,2}$}
\affiliation{$^{1}$Department of Mechanical and Materials
Engineering, Wright State University, Dayton, OH 45435}
\affiliation{$^{2}$National Research University Higher School of
Economics, Moscow 101000, Russia}
\date{\today}
\begin{abstract}

We investigate the effect of a single heat treatment cycle on the
potential energy states and mechanical properties of metallic
glasses using molecular dynamics simulations. We consider the
three-dimensional binary mixture, which was initially cooled with a
computationally slow rate from the liquid state to the solid phase
at a temperature well below the glass transition. It was found that
a cycle of heating and cooling can relocate the glass to either
rejuvenated or relaxed states, depending on the maximum temperature
and the loading period. Thus, the lowest potential energy is
attained after a cycle with the maximum temperature slightly below
the glass transition temperature and the effective cooling rate
slower than the initial annealing rate. In contrast, the degree of
rejuvenation increases when the maximum temperature becomes greater
than the glass transition temperature and the loading period is
sufficiently small. It was further shown that the variation of the
potential energy is inversely related to the dependence of the
elastic modulus and the yield stress as functions of the maximum
loading temperature. In addition, the heat treatment process causes
subtle changes in the shape of the radial distribution function of
small atoms.   These results are important for optimization of
thermal and mechanical processing of metallic glasses with
predetermined properties.

\vskip 0.5in


\end{abstract}

\maketitle

\section{Introduction}

The development of novel fabrication techniques for amorphous
materials, including metallic glasses and high entropy alloys, is
important for various industrial and biomedical
applications~\cite{LiAFM18,Branicio18}. The processing routes often
involve mechanical and thermal treatments of disordered alloys that
lead to changes in microstructure as well as mechanical and physical
properties~\cite{Greer16}. It is well realized by now that metallic
glasses relax to lower energy states during the aging process and
become more brittle, while rejuvenation and improved plasticity can
be achieved using a number of experimental techniques, such as shot
peening~\cite{Kerscher17}, cold rolling~\cite{Pelletier14}, high
pressure torsion~\cite{Ebner18}, ion irradiation~\cite{Gianola14},
and cryogenic thermal cycling~\cite{Ketov15,Guo18}.   In turn,
recent atomistic simulations have shown that periodically loaded
disordered materials evolve towards lower energy states at
sufficiently small strain amplitudes, whereas the yielding
transition and shear localization occur at large amplitudes after a
number of transient cycles~\cite{Lacks04,Priezjev13,Sastry13,
Reichhardt13,Priezjev14,Deng15,Priezjev16,Priezjev16a,Sastry17,
Priezjev17,Li17,Priezjev18,Priezjev18a,Alava18,NVP18strload,PriMakrho09}.
In addition, it was found that amorphous materials subjected to
multiple thermal cycles might either relax or rejuvenate depending
on the thermal amplitude, number of cycles, and processing
history~\cite{Priez18tcyc,Priez18T5000,Priez18T2000}. However, the
combined effect of thermal treatment and mechanical agitation on the
potential energy, structure, and mechanical properties of amorphous
alloys remains difficult to predict.

\vskip 0.05in

In the last few years, a number of studies have investigated the
aging and rejuvenation processes in amorphous materials subjected to
a heat treatment cycle using molecular dynamics simulations and
experimental measurements~\cite{Ogata15,Ogata16,EgamiFan17,Ogata17,
Barrat18,Maass18,Yang18}.   Most notably, it was demonstrated that
rejuvenated states can be attained by heating a sample above the
glass transition temperature, followed by isothermal annealing, and
then quenching with a rate higher than the initial cooling
rate~\cite{Ogata15}.   Moreover, the application of compressive
hydrostatic pressure during the quenching process was found to
promote thermal rejuvenation and improve plastic
performance~\cite{Ogata16,Yang18}.     More recently, it was shown
that the evolution of macroscopic state in disordered materials is
determined by the interplay between activation and relaxation on the
potential energy landscape, which helps to explain the thermal
hysteresis in cyclic thermal scanning~\cite{EgamiFan17}.   Using a
combination of calorimetry measurements and atomistic simulations, a
remarkable tenfold increase in stored energy was detected in rapidly
heated (above the glass transition) and cooled metallic
glasses~\cite{ Maass18}.  Despite considerable efforts, however, the
details of the heat treatment protocol, such as heating and cooling
rates, pressure, annealing temperature, and its effect on the
structure and properties of disordered solids remain not thoroughly
explored.

\vskip 0.05in

In this paper, we investigate how processing conditions influence
the potential energy states, structure, and mechanical properties of
amorphous materials subjected to a single heat treatment cycle.  We
consider a well-annealed binary mixture, which is linearly heated to
a maximum temperature in the neighborhood of the glass transition
temperature and cooled back to the glassy state. The simulations are
performed in a wide range of cycling periods and thermal amplitudes.
It will be shown that thermal cycling below the glass transition
relocates the system into states with lower potential energy, while
rejuvenated states can be accessed at sufficiently high
heating/cooling rates if the thermal amplitude is above the glass
transition temperature.

\vskip 0.05in

The rest of the paper is organized as follows. The molecular
dynamics simulation model and the thermal treatment protocol are
described in the next section. The variation of the potential energy
during thermal cycling and the resulted changes in the mechanical
properties as well as the structural analysis are presented in
section\,\ref{sec:Results}. The brief summary is given in the last
section.

\section{Details of molecular dynamics simulations}
\label{sec:MD_Model}

In the present work, we employed the binary mixture (80:20) model
introduced by Kob and Andersen (KA), which is similar to the
parametrization proposed by Weber and Stillinger to study the
amorphous metal alloy
$\text{Ni}_{80}\text{P}_{20}$~\cite{KobAnd95,Weber85}. In this
model, the interaction between atoms of different types, $A$ and
$B$, is strongly non-additive, thus preventing crystallization upon
cooling below the glass transition temperature. Any two atoms
$\alpha,\beta=A,B$ separated by a distance $r$ interact via the
Lennard-Jones (LJ) potential:
\begin{equation}
V_{\alpha\beta}(r)=4\,\varepsilon_{\alpha\beta}\,\Big[\Big(\frac{\sigma_{\alpha\beta}}{r}\Big)^{12}\!-
\Big(\frac{\sigma_{\alpha\beta}}{r}\Big)^{6}\,\Big],
\label{Eq:LJ_KA}
\end{equation}
with the interaction parameters: $\varepsilon_{AA}=1.0$,
$\varepsilon_{AB}=1.5$, $\varepsilon_{BB}=0.5$, $\sigma_{AA}=1.0$,
$\sigma_{AB}=0.8$, $\sigma_{BB}=0.88$, and
$m_{A}=m_{B}$~\cite{KobAnd95}. The system consists of $N=60\,000$
atoms, and thus the cutoff radius of the LJ potential is set to
$r_{c,\,\alpha\beta}=2.5\,\sigma_{\alpha\beta}$ to improve the
computational efficiency. Throughout the study, the physical
quantities are expressed in terms of the LJ units of length, mass,
energy, and time: $\sigma=\sigma_{AA}$, $m=m_{A}$,
$\varepsilon=\varepsilon_{AA}$, and
$\tau=\sigma\sqrt{m/\varepsilon}$. The integration time step was
fixed to $\triangle t_{MD}=0.005\,\tau$ in the LAMMPS
software~\cite{Lammps}.

\vskip 0.05in

% equilibration and temperature protocol

The binary mixture was first thoroughly equilibrated in the liquid
state at the temperature $T_{LJ}=1.0\,\varepsilon/k_B$ and zero
pressure. In what follows, $k_B$ denotes the Boltzmann constant. The
temperature was regulated via the Nos\'{e}-Hoover
thermostat~\cite{Allen87}. The simulations were carried out in the
$NPT$ ensemble, and the periodic boundary conditions were imposed
along three dimensions~\cite{Allen87}. After equilibration, the
system was cooled with computationally slow rate of
$10^{-5}\varepsilon/k_{B}\tau$ at zero pressure to the temperature
of $0.01\,\varepsilon/k_B$.  Next, the glass was subjected to one
cycle of heating to the maximum temperature and then cooling back to
$T_{LJ}=0.01\,\varepsilon/k_B$ at $P=0$ during the time period $T$.
To avoid confusion, we denote the maximum temperature during the
thermal cycle by $T_{LJ}^M$, temperature of the system by $T_{LJ}$,
and the oscillation period by $T$. During the production runs, the
potential energy, temperature, system dimensions, stresses and
atomic configurations were saved for post-processing and
visualization.  In order to examine the changes in mechanical
properties, the binary glass was strained before and after the
thermal cycle along the $\hat{x}$ direction with the strain rate
$\dot{\varepsilon}_{xx}=10^{-5}\,\tau^{-1}$ at
$T_{LJ}=0.01\,\varepsilon/k_B$ and $P=0$. The stress-strain curves
were used to estimate the elastic modulus and the peak value of the
stress overshoot for different values of the parameters $T_{LJ}^M$
and $T$. The data were averaged over ten independent realizations of
disorder. The MD simulations of the thermal loading process with
large periods required about 2000 processors.

\section{Results}
\label{sec:Results}

% intro and brief review

Upon cooling a glass-former from the liquid state to the solid
phase, its structure remains amorphous but the molecular motion
dramatically slows down~\cite{Debenedetti01}.   Moreover, with
decreasing cooling rate, the glass transition temperature is reduced
and the potential energy of the disordered solid becomes lower, as
the system has more time to explore different configurations and
find a deeper minimum~\cite{Debenedetti01}. It is also well known
that slowly cooled and/or aged glasses exhibit higher yield stress,
followed by the formation of sharply localized shear
bands~\cite{Greer16}.   By contrast, mechanically or thermally
rejuvenated glasses can undergo a more homogeneous plastic
deformation and thus they are generally less brittle~\cite{Greer16}.
It was recently shown that the level of rejuvenation can be
controlled by applying thermal treatment, where the glass is first
heated into the liquid state, annealed during a certain time
interval, and then rapidly quenched to the solid
phase~\cite{Ogata15}.

% In the present study, the effect of a simple heating and cooling
% cycle is considered in a wide range of

% Aside from mechanical processing, review cooling rates protocol
% thermal treatment processing

\vskip 0.05in

% comment on Tg

\begin{figure}[t]
\includegraphics[width=12.0cm,angle=0]{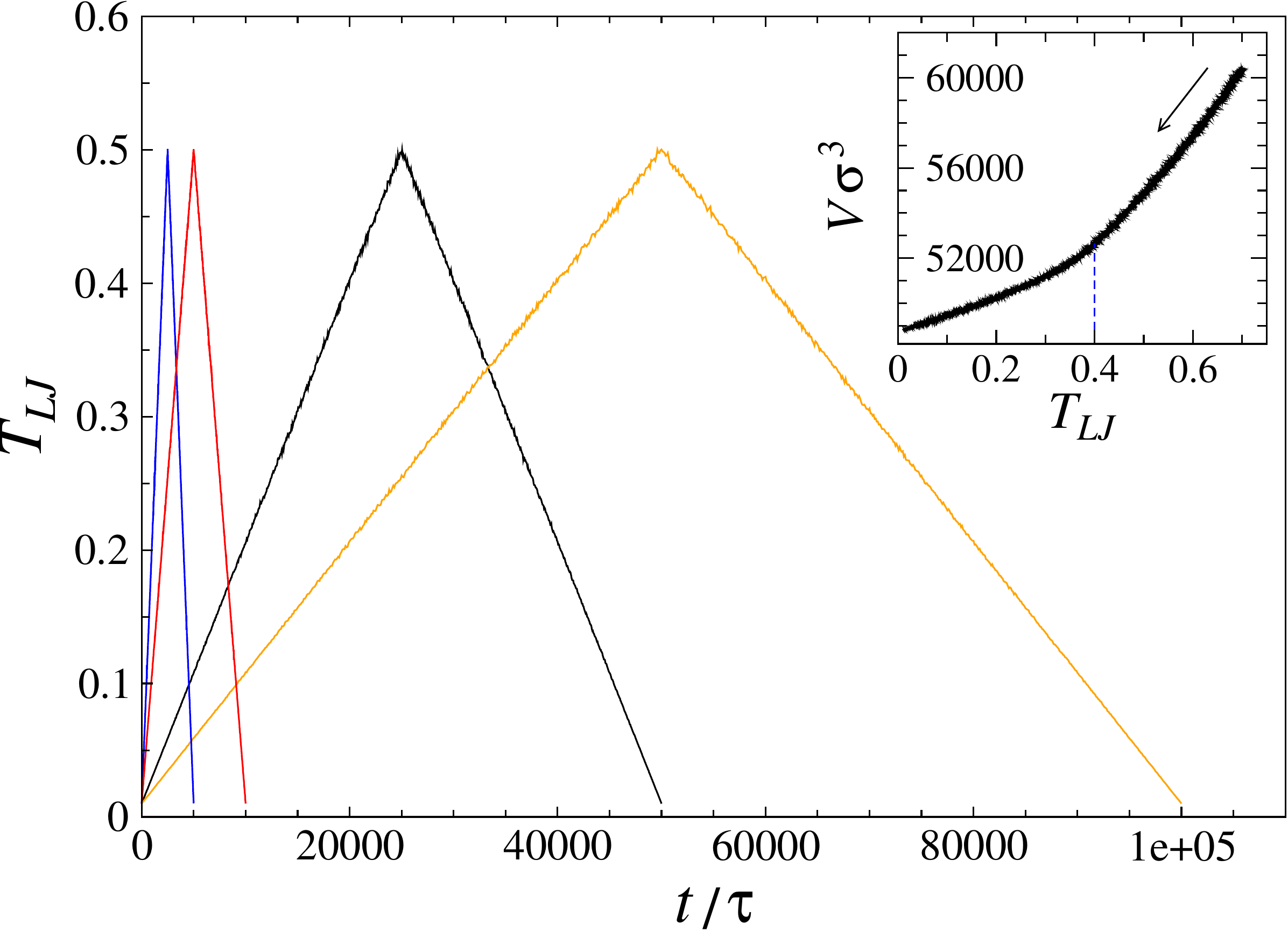}
\caption{(Color online) The temperature profiles $T_{LJ}$ (in units
of $\varepsilon/k_B$) measured during one cycle of heating and
cooling with the periods $T=5000\,\tau$ (red), $T=10\,000\,\tau$
(blue), $T=50\,000\,\tau$ (black), and $T=100\,000\,\tau$ (orange).
The inset shows the variation of volume as a function of temperature
during initial cooling with the rate of
$10^{-5}\varepsilon/k_{B}\tau$ to the temperature
$T_{LJ}=0.01\,\varepsilon/k_B$. The vertical dashed line indicates
the glass transition temperature of about $0.40\,\varepsilon/k_B$ at
the density $\rho\approx1.14\,\sigma^{-3}$. }
\label{fig:temp_control}
\end{figure}

In our study, the disordered solids were initially prepared by
cooling with the rate of $10^{-5}\varepsilon/k_{B}\tau$ to the
temperature $T_{LJ}=0.01\,\varepsilon/k_B$ at zero pressure. As an
example, the dependence of volume as a function of temperature for
one sample is presented in the inset to
Fig.\,\ref{fig:temp_control}. The intersection of linearly
extrapolated low (glass phase) and high (liquid phase) temperature
regions provides an estimate of the glass transition temperature of
about $0.40\,\varepsilon/k_B$. Correspondingly, the average glass
density at this temperature is $\rho\approx1.14\,\sigma^{-3}$. Note
that the glass transition temperature at zero pressure is lower than
the mode-coupling critical temperature $T_c=0.435\,\varepsilon/k_B$
at the density $\rho=1.2\,\sigma^{-3}$, which was determined
numerically by fitting the diffusion coefficient to the power-law
function of temperature at constant volume~\cite{KobAnd95}.

\vskip 0.05in

% thermal cycle protocol

The examples of temperature profiles measured in MD simulations
during one cycle of heating and cooling are presented in
Fig.\,\ref{fig:temp_control} for the periods $T=5000\,\tau$,
$10\,000\,\tau$, $50\,000\,\tau$ and $100\,000\,\tau$. Note that the
effective heating and cooling rates are the same and equal to
approximately $2\,T_{LJ}^M/T$.   Thus, the rate varies from
$0.8\times10^{-4}\varepsilon/k_{B}\tau$ at
$T_{LJ}^M=0.20\,\varepsilon/k_B$ to
$2.8\times10^{-4}\varepsilon/k_{B}\tau$ at
$T_{LJ}^M=0.70\,\varepsilon/k_B$ for the smallest period
$T=5000\,\tau$, while it ranges from
$0.8\times10^{-6}\varepsilon/k_{B}\tau$ at
$T_{LJ}^M=0.20\,\varepsilon/k_B$ to
$2.8\times10^{-6}\varepsilon/k_{B}\tau$ at
$T_{LJ}^M=0.70\,\varepsilon/k_B$ for the largest period
$T=500\,000\,\tau$. In the following analysis, the change in the
potential energy due to thermal cycling will be correlated with the
relative difference between these rates and the initial cooling rate
of $10^{-5}\varepsilon/k_{B}\tau$.

\vskip 0.05in

% potential energy versus temperature

\begin{figure}[t]
\includegraphics[width=12.0cm,angle=0]{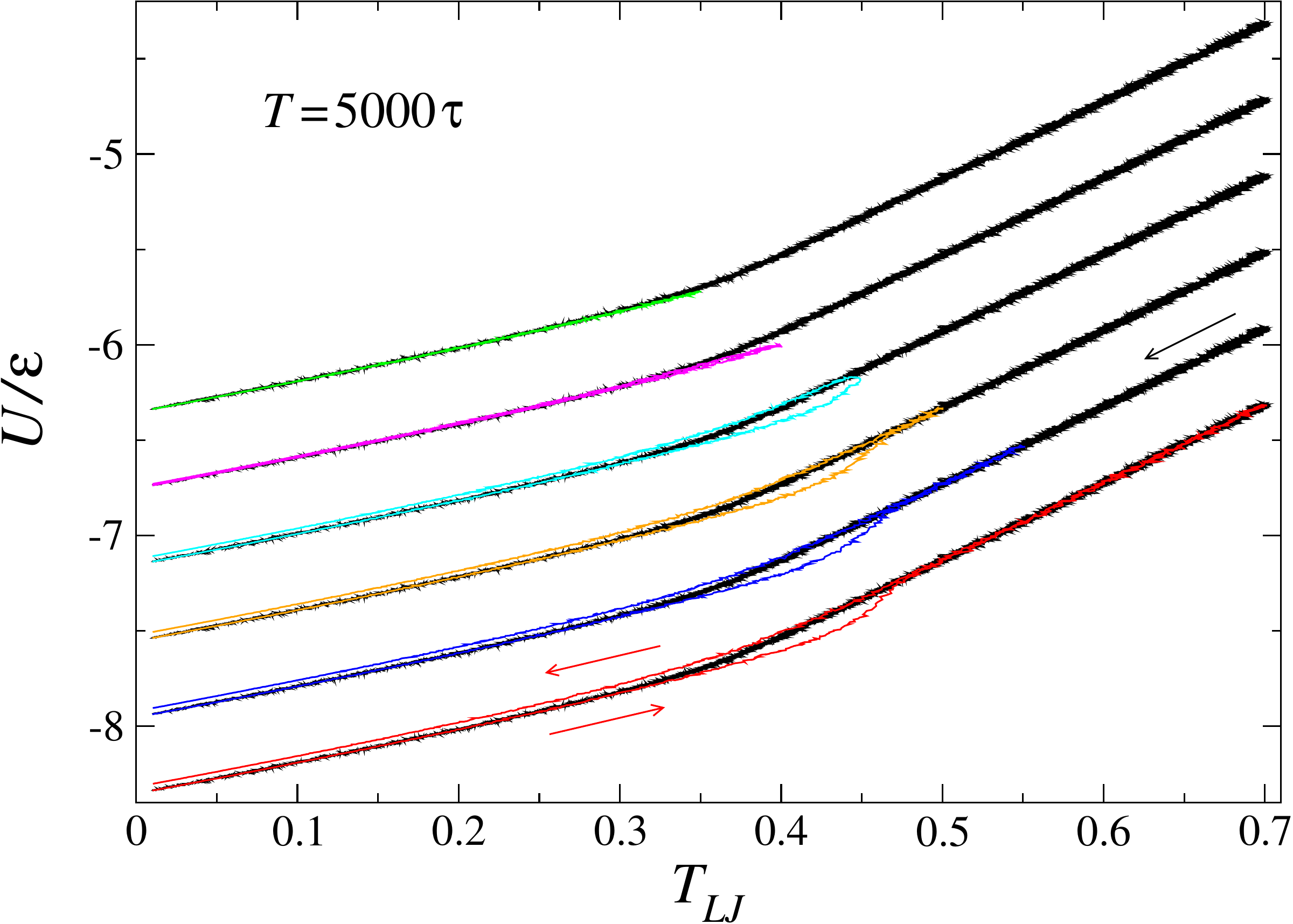}
\caption{(Color online) The variation of the potential energy during
one cycle with the period $T=5000\,\tau$ and the maximum temperature
$T_{LJ}^M=0.35\,\varepsilon/k_B$ (green), $0.40\,\varepsilon/k_B$
(magenta), $0.45\,\varepsilon/k_B$ (cyan), $0.50\,\varepsilon/k_B$
(orange), and $0.55\,\varepsilon/k_B$ (blue), and
$0.70\,\varepsilon/k_B$ (red).  The black curves indicate the
potential energy during cooling with the rate
$10^{-5}\varepsilon/k_{B}\tau$. The data for $0.35\,\varepsilon/k_B
\leqslant T_{LJ}^M \leqslant 0.55\,\varepsilon/k_B$ and the
corresponding black curves are displaced vertically for clarity. }
\label{fig:poten_TLJ_p5000}
\end{figure}

\begin{figure}[t]
\includegraphics[width=12.0cm,angle=0]{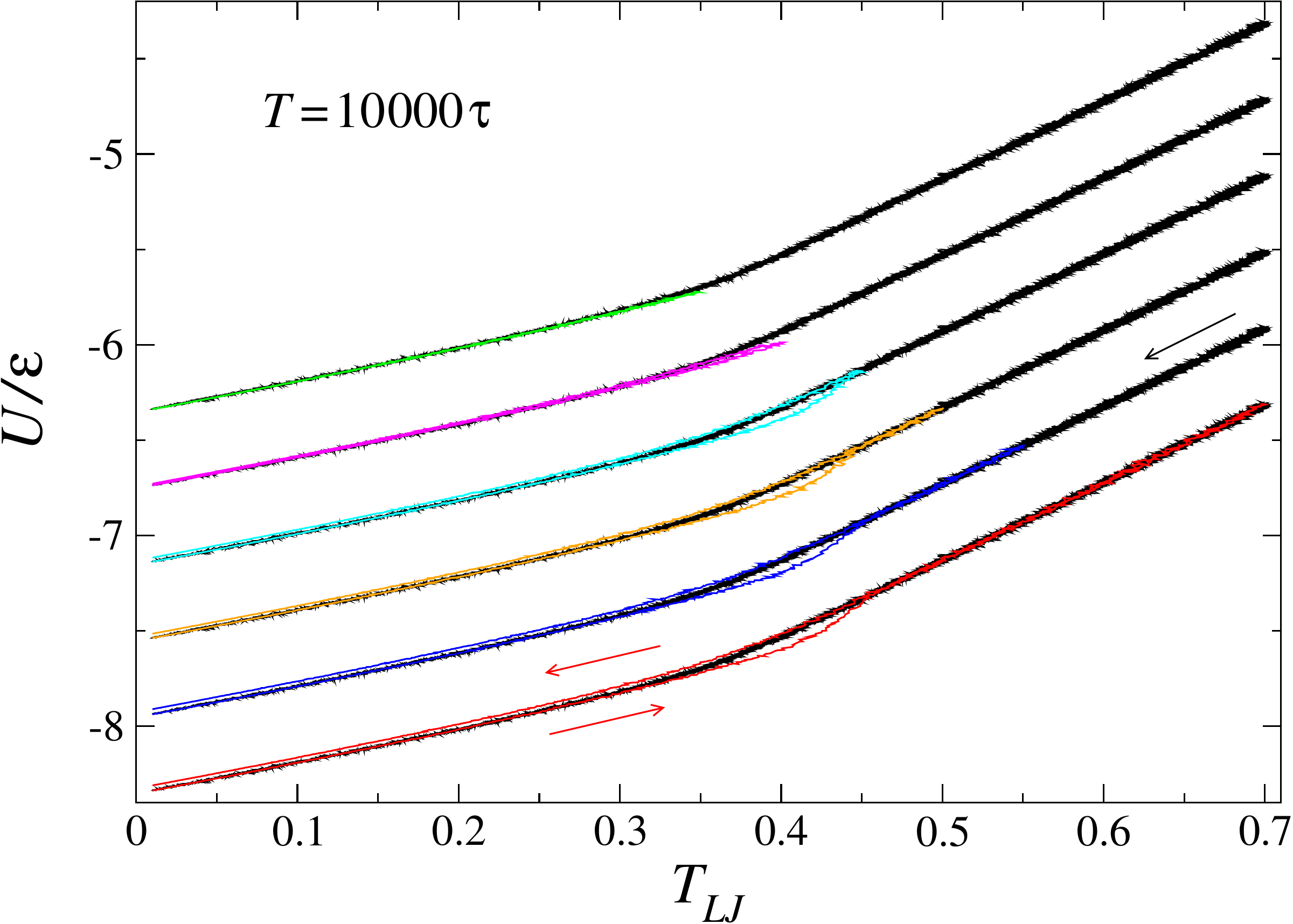}
\caption{(Color online) The potential energy during heating and
cooling with the period $T=10\,000\,\tau$ and the maximum
temperature $T_{LJ}^M=0.35\,\varepsilon/k_B$ (green),
$0.40\,\varepsilon/k_B$ (magenta), $0.45\,\varepsilon/k_B$ (cyan),
$0.50\,\varepsilon/k_B$ (orange), and $0.55\,\varepsilon/k_B$
(blue), and $0.70\,\varepsilon/k_B$ (red).   The black curves denote
the potential energy during initial cooling with the rate
$10^{-5}\varepsilon/k_{B}\tau$. All data, except for the two lowest
curves (black and red), are displaced upwards for clarity. }
\label{fig:poten_TLJ_p10000}
\end{figure}

\begin{figure}[t]
\includegraphics[width=12.0cm,angle=0]{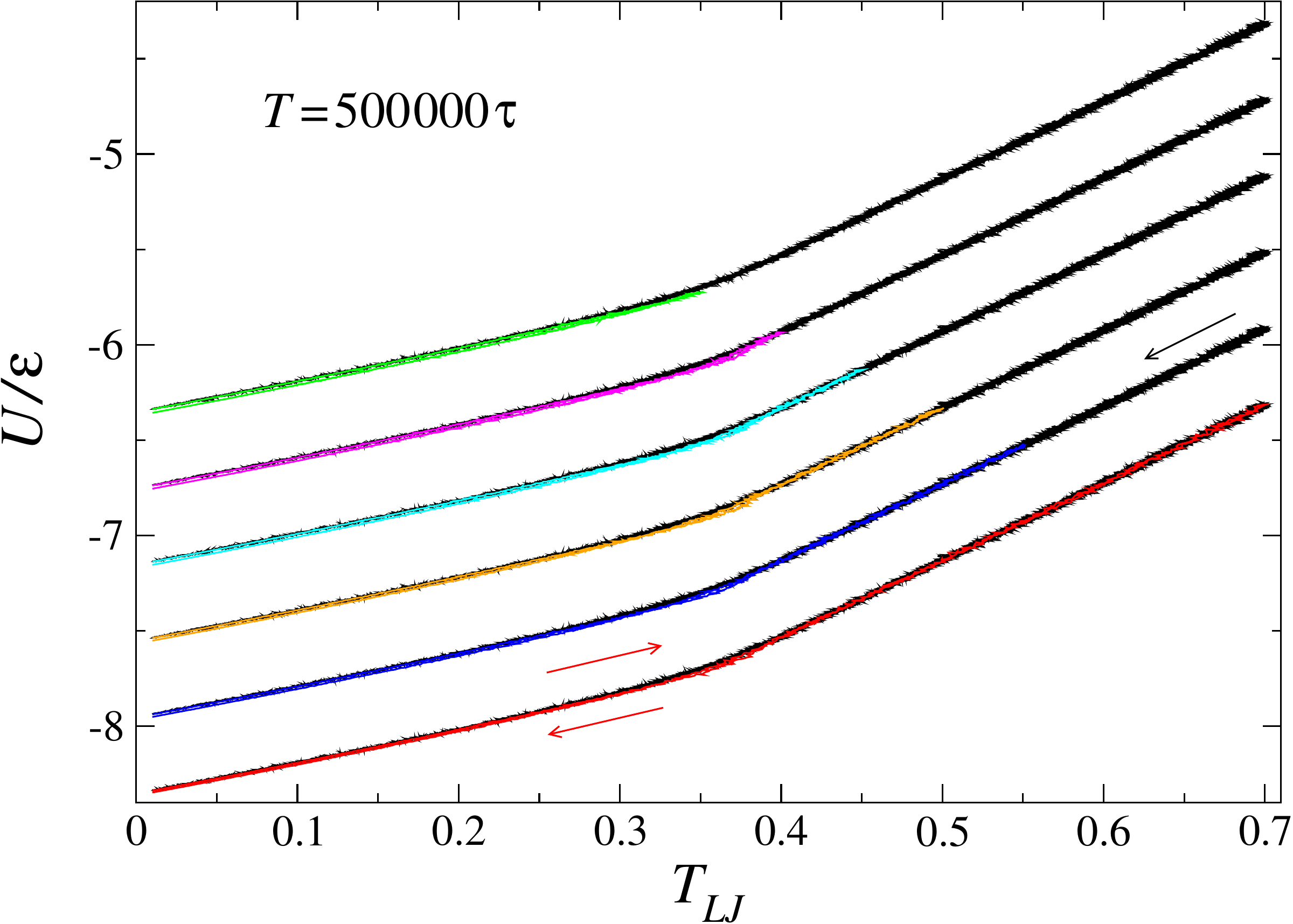}
\caption{(Color online) The dependence of the potential energy
during the thermal cycle with the period $T=500\,000\,\tau$ and the
maximum temperature $T_{LJ}^M=0.35\,\varepsilon/k_B$ (green),
$0.40\,\varepsilon/k_B$ (magenta), $0.45\,\varepsilon/k_B$ (cyan),
$0.50\,\varepsilon/k_B$ (orange), and $0.55\,\varepsilon/k_B$
(blue), and $0.70\,\varepsilon/k_B$ (red).  The potential energy
during cooling with the rate $10^{-5}\varepsilon/k_{B}\tau$ is
indicated by the black curves. Note that the data for
$0.35\,\varepsilon/k_B \leqslant T_{LJ}^M \leqslant
0.55\,\varepsilon/k_B$ are displaced for clarity. }
\label{fig:poten_TLJ_p500000}
\end{figure}

The variation of the potential energy per atom during the heating
and cooling cycle is presented in
Figs.\,\ref{fig:poten_TLJ_p5000}--\ref{fig:poten_TLJ_p500000} for
cycling periods $T=5000\,\tau$, $10\,000\,\tau$, and
$500\,000\,\tau$ for selected values of the maximum temperature. For
reference, the potential energy upon cooling from
$T_{LJ}=0.7\,\varepsilon/k_B$ to $T_{LJ}=0.01\,\varepsilon/k_B$ with
the rate $10^{-5}\varepsilon/k_{B}\tau$ is shown by the black
curves.     The data denoted by the black curves are the same, and,
except for the lowest one, the curves are displaced vertically for
clarity. The directions of the temperature variation in each case
are indicated by the corresponding arrows.  We also comment that
extrapolation of $U(T_{LJ})$ from the glass and liquid regions
provides an estimate of the glass transition temperature
$T_g\approx0.35\,\varepsilon/k_B$, which is lower than the value of
$0.40\,\varepsilon/k_B$ obtained from the data $V(T_{LJ})$ shown in
the inset to Fig.\,\ref{fig:temp_control}.

\vskip 0.05in

% potential energy versus temperature; cont

As shown in Fig.\,\ref{fig:poten_TLJ_p5000}, the potential energy
during thermal loading with the smallest period $T=5000\,\tau$ and
$T_{LJ}^M=0.35\,\varepsilon/k_B$ essentially coincides with the data
obtained during initial cooling with the rate
$10^{-5}\varepsilon/k_{B}\tau$, and thus the potential energy before
and after the thermal cycle remains nearly the same.  Note that in
the case $T_{LJ}^M=0.45\,\varepsilon/k_B$, the heating rate
$1.8\times10^{-4}\varepsilon/k_{B}\tau$ is greater than
$10^{-5}\varepsilon/k_{B}\tau$, and the system enters the region
above the glass transition with the potential energy smaller than
$U$ during the initial cooling (denoted by the black curve). When
$T_{LJ}\gtrsim0.37\,\varepsilon/k_B$, the system is first
rejuvenated and then cooled down with the rate of
$1.8\times10^{-4}\varepsilon/k_{B}\tau$ (greater than
$10^{-5}\varepsilon/k_{B}\tau$), resulting in higher energy states
(see the cyan curve in Fig.\,\ref{fig:poten_TLJ_p5000}). Finally,
during the thermal treatment with the maximum temperatures
$T_{LJ}^M\geqslant0.50\,\varepsilon/k_B$, the potential energy
follows the black curve up to $T_{LJ}\approx0.35\,\varepsilon/k_B$
and above $T_{LJ}\approx0.47\,\varepsilon/k_B$ (see
Fig.\,\ref{fig:poten_TLJ_p5000}).   In these cases, the change in
the potential energy due to thermal loading is simply determined by
the effective cooling rate $2\,T_{LJ}^M/T$; \textit{i.e.}, the
potential energy at the end of the cycle becomes higher with
increasing $T_{LJ}^M$.

\vskip 0.05in

% potential energy versus temperature; cont

The same trends can be observed in Fig.\,\ref{fig:poten_TLJ_p10000}
for thermal cycling with the larger period $T=10\,000\,\tau$, except
that the area of the hysteresis loops near $T_g$ becomes smaller and
the potential energy difference after a full cycle is reduced due to
the lower cooling rate $2\,T_{LJ}^M/T$.   By contrast, a
qualitatively different behavior occurs for the largest period
$T=500\,000\,\tau$, as shown in Fig.\,\ref{fig:poten_TLJ_p500000}.
In this case, the heating and cooling rates $2\,T_{LJ}^M/T$ are
smaller than the initial cooling rate
$10^{-5}\varepsilon/k_{B}\tau$.   During the thermal loading with
$T_{LJ}^M=0.35\,\varepsilon/k_B$, the system remains in the glass
phase, and the aging is accelerated when the temperature is close to
$T_g$, resulting in a noticeable decrease of the potential energy at
the end of the cycle (see Fig.\,\ref{fig:poten_TLJ_p500000}).
Furthermore, when $T_{LJ}^M>T_g$, the system enters the liquid phase
and then cooled with the rates $2\,T_{LJ}^M/T$ slower than
$10^{-5}\varepsilon/k_{B}\tau$, and thus the potential energy after
the thermal cycle is also reduced. This behavior is consistent with
the simulation results of thermal processing of Cu-Zr amorphous
alloys~\cite{Ogata15,EgamiFan17}.

\vskip 0.05in

% summary potential energy for thermal cycling

\begin{figure}[t]
\includegraphics[width=12.0cm,angle=0]{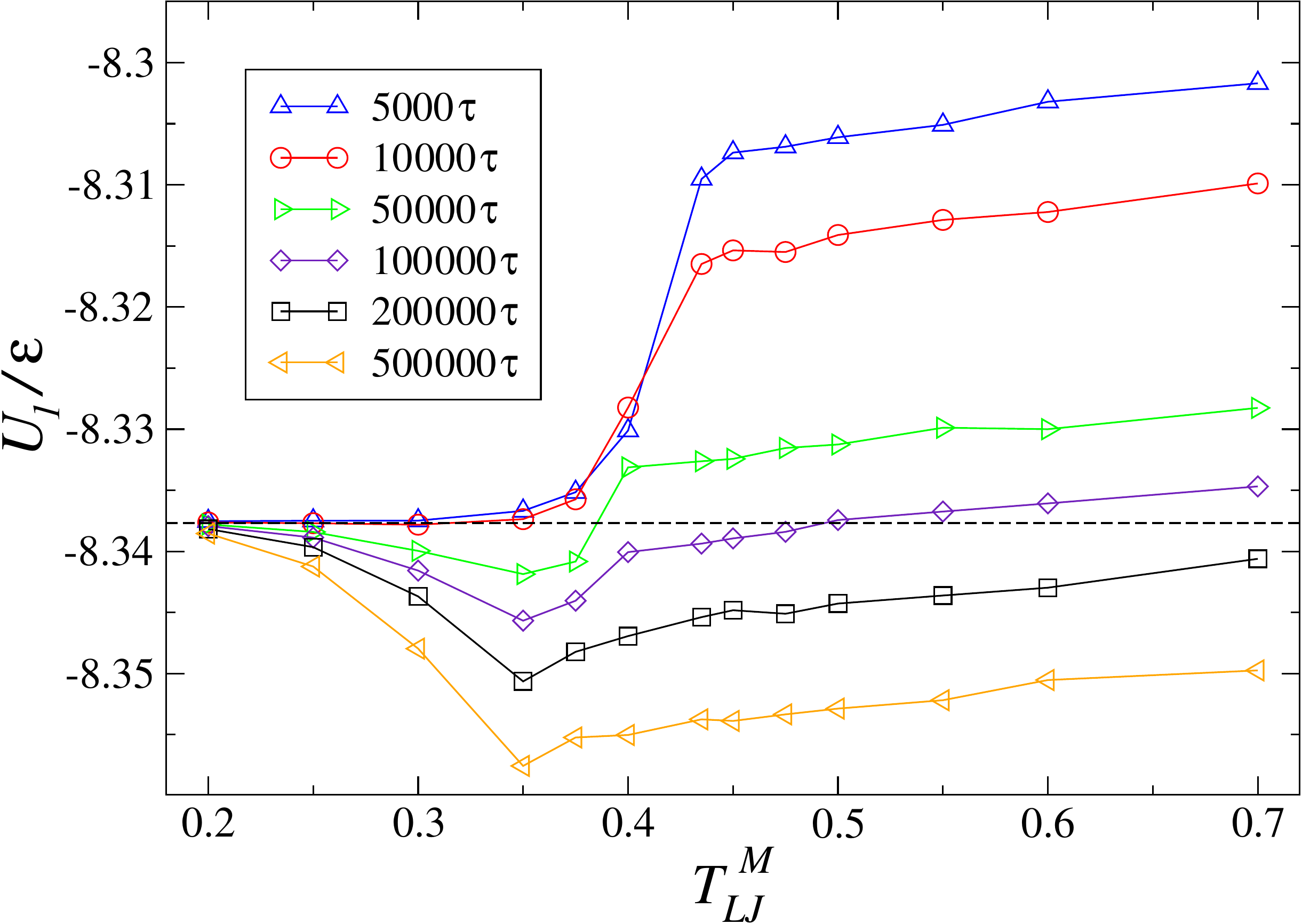}
\caption{(Color online) The potential energy after the thermal
cycle, $U_{1}/\varepsilon$, as a function of the thermal amplitude
$T_{LJ}^{M}$ (in units of $\varepsilon/k_B$) for the oscillation
periods $T=5000\,\tau$ ($\vartriangle$), $10\,000\,\tau$ ($\circ$),
$50\,000\,\tau$ ($\vartriangleright$), $100\,000\,\tau$
($\diamond$), $200\,000\,\tau$ ($\square$), and $500\,000\,\tau$
($\vartriangleleft$). The horizontal dashed line denotes the
potential energy level at the beginning of the thermal cycle. }
\label{fig:U1_delT}
\end{figure}

The summary of the data for the potential energy after the thermal
cycle, $U_1/\varepsilon$, is presented in Fig.\,\ref{fig:U1_delT}
for the indicated loading periods. For comparison, the potential
energy level after the initial cooling with the rate of
$10^{-5}\varepsilon/k_{B}\tau$, but before the thermal treatment, is
denoted by the horizontal dashed line.   Several features are
noteworthy.  It can be seen in Fig.\,\ref{fig:U1_delT} that the
potential energy after the thermal cycle with the maximum
temperature $T_{LJ}^M \lesssim0.35\,\varepsilon/k_B$ remains nearly
unchanged for small periods, $T=5000\,\tau$ and $T=10\,000\,\tau$,
while $U_1$ is reduced for larger periods.   With increasing loading
period, the energy difference increases, and the lowest value
$U_1\approx-8.358\,\varepsilon$ is attained at
$T_{LJ}^M=0.35\,\varepsilon/k_B$ when $T=500\,000\,\tau$.
Interestingly, it was recently found that approximately the same
value of the potential energy $U\approx-8.356\,\varepsilon$ was
obtained after 100 thermal cycles with the period $T=5000\,\tau$ and
maximum temperature $T_{LJ}^M=0.35\,\varepsilon/k_B$ for the glass
prepared with the cooling rate
$10^{-5}\varepsilon/k_{B}\tau$~\cite{Priez18T5000}. These results
suggest that the aging process, which is accelerated in the vicinity
of $T_g$, leads to the same energy decrease regardless whether the
waiting time interval is continuous or discrete.

\vskip 0.05in

It should also be commented that the possibility of attaining higher
energy states upon cryogenic thermal cycling (well below $T_g$) was
recently discussed by Shang \textit{et al.}~\cite{Barrat18}.   In
particular, it was concluded that internal stresses due to
heterogeneity in the local thermal expansion can induce local shear
transformations, provided that the system size is sufficiently
large~\cite{Barrat18}. The efficiency of thermal rejuvenation
generally increases with the number of thermal cycles and becomes
relatively large when the system dimensions exceed about
10\,nm~\cite{Barrat18}.  In the present study, however, the
appearance of rejuvenated states in Fig.\,\ref{fig:U1_delT} for
small periods $T=5000\,\tau$ and $10\,000\,\tau$ when
$0.35\,\varepsilon/k_B\leqslant T_{LJ}^M
\leqslant0.40\,\varepsilon/k_B$ is related to the small hysteresis
near $T_g$, as shown in Figs.\,\ref{fig:poten_TLJ_p5000} and
\ref{fig:poten_TLJ_p10000}.  In other words, the aging effects are
negligible when $T\lesssim10\,000\,\tau$, and the system approaches
the glass transition temperature, becomes slightly rejuvenated, and
then quenched with a high rate into the glass phase, resulting in
higher potential energy states.

\vskip 0.05in

As illustrated in Fig.\,\ref{fig:U1_delT}, the thermal loading
process with the maximum temperature
$T_{LJ}^M\gtrsim0.4\,\varepsilon/k_B$ can lead to either rejuvenated
or relaxed states depending on the effective cooling rate
$2\,T_{LJ}^M/T$. Thus, the energy difference is positive when
$T\geqslant100\,000\,\tau$, and it is mainly determined by the time
interval when $T_{LJ}$ is above the glass transition temperature,
$T\,(T_{LJ}^M-T_g)/T_{LJ}^M$, as well as the ratio between the rates
$2\,T_{LJ}^M/T$ and $10^{-5}\varepsilon/k_{B}\tau$. Note that the
potential energy before and after thermal loading with
$T_{LJ}^M=0.50\,\varepsilon/k_B$ and $T=100\,000\,\tau$ is nearly
the same, since the cooling rate is essentially equal to the initial
cooling rate $10^{-5}\varepsilon/k_{B}\tau$. As is evident from
Fig.\,\ref{fig:U1_delT}, $U_1$ reaches a quasi-plateau for all
cycling periods when $T_{LJ}^M\gtrsim0.45\,\varepsilon/k_B$, and its
slope is directly related to the increase in the cooling rate
$2\,T_{LJ}^M/T$.

% thus above rejuvenation, below aging

\vskip 0.05in

% structural analysis

\begin{figure}[t]
\includegraphics[width=12.0cm,angle=0]{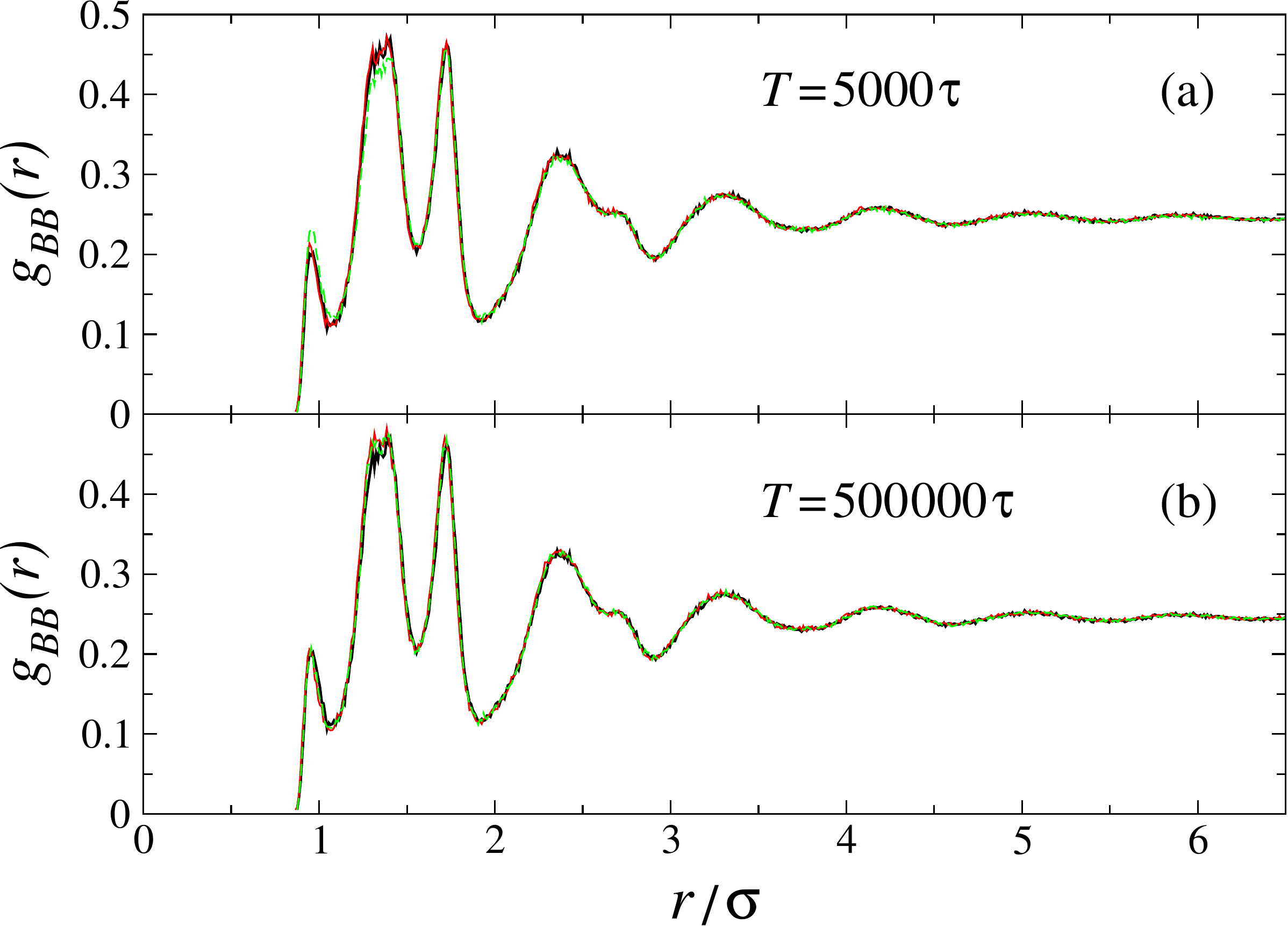}
\caption{(Color online) The radial distribution function,
$g_{BB}(r)$, for binary glasses at $T_{LJ}=0.01\,\varepsilon/k_B$
after thermal loading with periods (a) $T=5000\,\tau$ and (b)
$T=500\,000\,\tau$. The maximum temperatures of the thermal
treatment are $T_{LJ}^M=0.35\,\varepsilon/k_B$ (solid red) and
$T_{LJ}^M=0.70\,\varepsilon/k_B$ (dashed green).  The data after the
initial cooling with the rate $10^{-5}\varepsilon/k_{B}\tau$ are
denoted by the solid black curves. }
\label{fig:grBB}
\end{figure}

Although the atomic structure before and after the thermal loading
remains amorphous, some subtle differences can be detected by
examining the radial distribution function.   In the previous MD
study of the KA binary mixture, it was shown that one of the most
sensitive measures of structural changes upon temperature variation
is the distribution function of small atoms of type
$B$~\cite{Stillinger00}. In Fig.\,\ref{fig:grBB}, we plot the
averaged distribution function of $B\!-\!B$ atoms $g_{BB}(r)$ for
the smallest $T=5000\,\tau$ and largest $T=500\,000\,\tau$ periods.
Two values of the maximum temperature were chosen,
$T_{LJ}^M=0.35\,\varepsilon/k_B$ and $0.70\,\varepsilon/k_B$, at
which the aging and rejuvenation effects are most pronounced. For
reference, the data before the thermal loading are also included in
Fig.\,\ref{fig:grBB}. It can be seen in Fig.\,\ref{fig:grBB}\,(a)
that the most rejuvenated state at $T=5000\,\tau$ and
$T_{LJ}^M=0.70\,\varepsilon/k_B$ is characterized by a slightly
higher value of the peak height at $r\approx1.0\,\sigma$, while the
second peak becomes lower.   In contrast, the structure of the aged
glass loaded during $T=500\,000\,\tau$ at
$T_{LJ}^M=0.35\,\varepsilon/k_B$ has a slightly higher second peak,
as shown in Fig.\,\ref{fig:grBB}\,(b).  It should be also mentioned
that we did not detect a pronounced icosahedral order either before
or after thermal loading, and, therefore, it is not reported here.
It was found, however, that aging increases and rejuvenation
decreases the icosahedral short-range order in Zr-based metallic
glasses~\cite{Ogata15}.

\vskip 0.05in

% mechanical properties

\begin{figure}[t]
\includegraphics[width=12.0cm,angle=0]{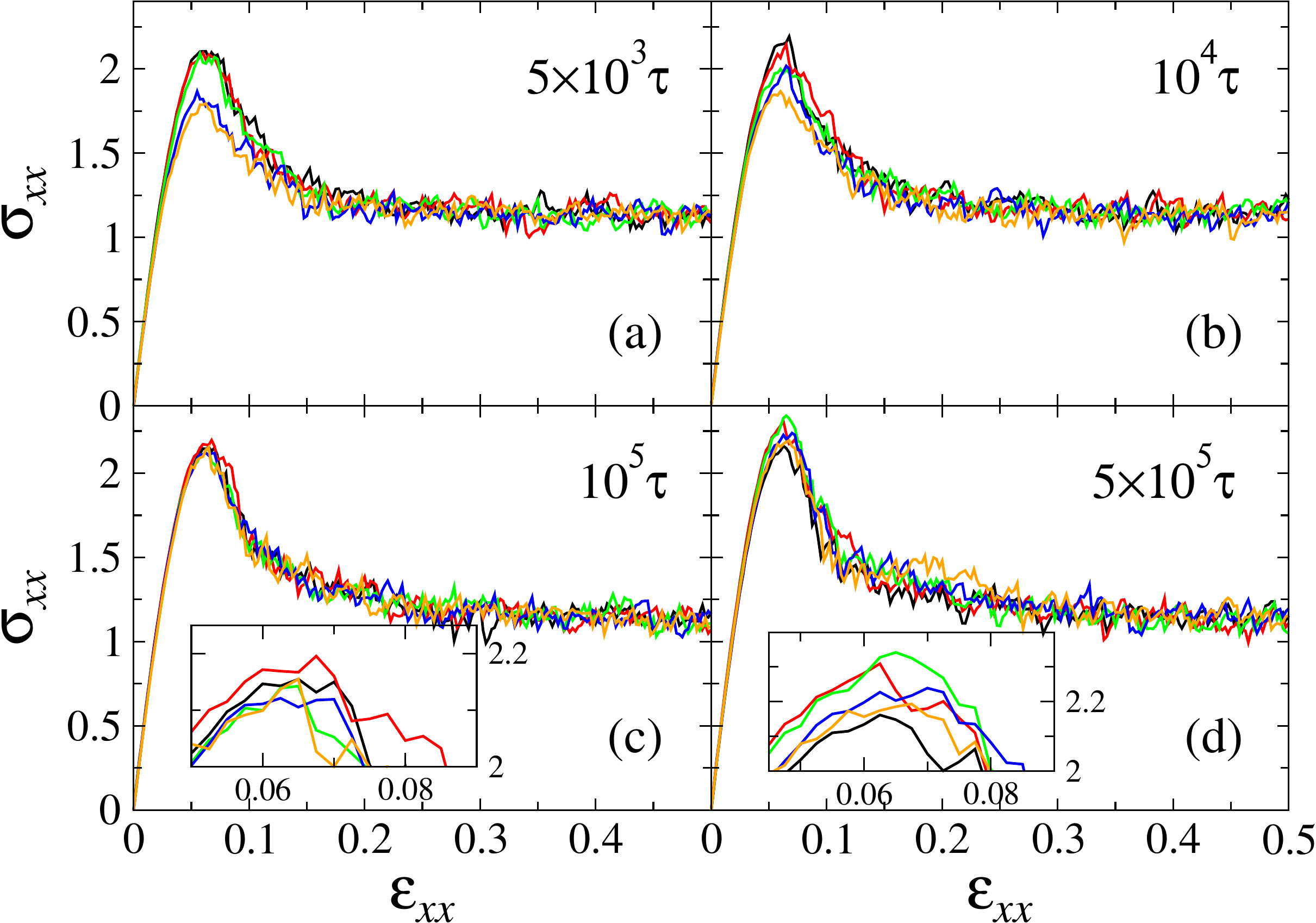}
\caption{(Color online) The tensile stress $\sigma_{xx}$ (in units
of $\varepsilon\sigma^{-3}$) as a function of strain,
$\varepsilon_{xx}$, for binary glasses after one thermal cycle with
the period (a) $T=5000\,\tau$, (b) $T=10\,000\,\tau$, (c)
$T=100\,000\,\tau$, and (d) $T=500\,000\,\tau$. The maximum
temperature is $T_{LJ}^{M}=0.20\,\varepsilon/k_B$ (black),
$0.375\,\varepsilon/k_B$ (red), $0.40\,\varepsilon/k_B$ (green),
$0.435\,\varepsilon/k_B$ (blue), and $0.7\,\varepsilon/k_B$
(orange). The insets show the enlarged view of the same data. The
rate of strain is $\dot{\varepsilon}_{xx}=10^{-5}\,\tau^{-1}$ and
the temperature is $T_{LJ}=0.01\,\varepsilon/k_B$. }
\label{fig:stress_strain}
\end{figure}

\begin{figure}[t]
\includegraphics[width=12.cm,angle=0]{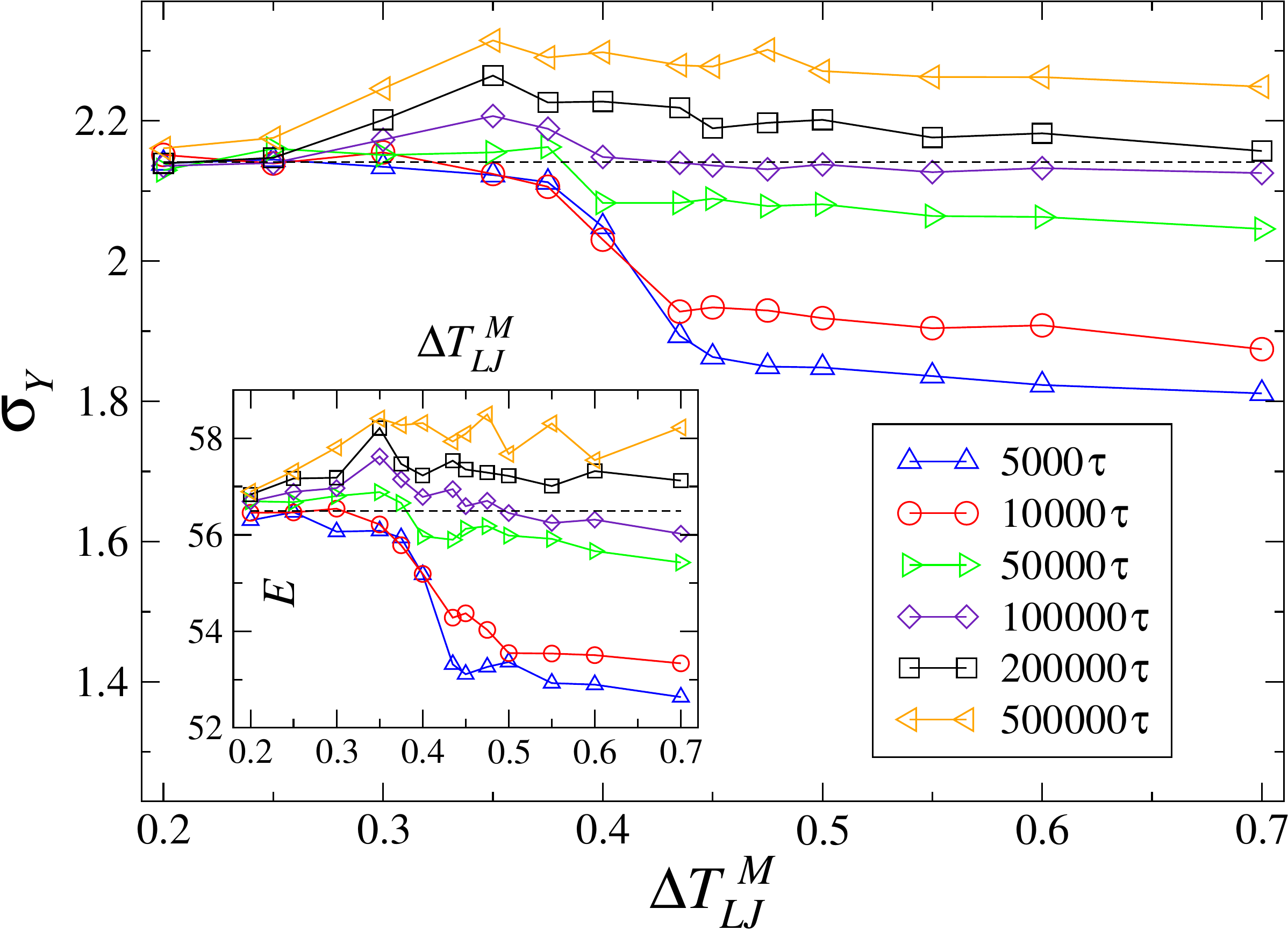}
\caption{(Color online) The peak value of the stress overshoot
$\sigma_Y$ (in units of $\varepsilon\sigma^{-3}$) as a function of
the maximum temperature of the loading cycle. The inset shows the
elastic modulus $E$ (in units of $\varepsilon\sigma^{-3}$) versus
$T_{LJ}^{M}$ (in units of $\varepsilon/k_B$). The mechanical
properties were probed after the thermal cycle with periods
$T=5000\,\tau$ ($\vartriangle$), $10\,000\,\tau$ ($\circ$),
$50\,000\,\tau$ ($\vartriangleright$), $100\,000\,\tau$
($\diamond$), $200\,000\,\tau$ ($\square$), and $500\,000\,\tau$
($\vartriangleleft$). The data before the loading cycle are denoted
by the horizontal dashed lines.}
\label{fig:yield_stress_E}
\end{figure}

The representative stress-strain curves after the thermal loading
process are presented in Fig.\,\ref{fig:stress_strain} for different
cycling periods.  The samples were uniaxially strained with a
constant rate $\dot{\varepsilon}_{xx}=10^{-5}\,\tau^{-1}$ at
$T_{LJ}=0.01\,\varepsilon/k_B$ and $P=0$. In all cases, the tensile
stress exhibits a distinct peak, followed by a steady flow at higher
strain.  It can be observed in Fig.\,\ref{fig:stress_strain}\,(a)
that the largest decrease in the peak height occurs after loading
with the period $T=5000\,\tau$ when the maximum temperature varies
from $T_{LJ}^M=0.2\,\varepsilon/k_B$ to $0.7\,\varepsilon/k_B$. This
trend is inversely correlated with the behavior of $U_1(T_{LJ})$ for
$T=5000\,\tau$ shown in Fig.\,\ref{fig:U1_delT}.   On the other
hand, as $T_{LJ}^M$ increases, the stress overshoot becomes larger
than its value for $T_{LJ}^M=0.2\,\varepsilon/k_B$ for the largest
period $T=500\,000\,\tau$, see Fig.\,\ref{fig:stress_strain}\,(d).
Next, the elastic modulus was computed from the linear fit of the
stress-strain curves for $\varepsilon_{xx}\leqslant0.01$ and then
averaged over 10 samples for each $T_{LJ}^M$ and $T$.  The data for
the stress overshoot, $\sigma_{Y}$, and the elastic modulus, $E$,
are plotted in Fig.\,\ref{fig:yield_stress_E}.   Overall, the
variation of mechanical properties of thermally loaded samples,
compared to untreated glasses, correlates well with the change in
the potential energy reported in Fig.\,\ref{fig:U1_delT}. In other
words, rejuvenated glasses are characterized by reduced values of
$\sigma_{Y}$ and $E$, while the opposite trend is observed for aged
samples.   The maximum decrease in $\sigma_{Y}$ and $E$ of about
10\% is detected for the smallest period $T=5000\,\tau$ and
$T_{LJ}^M \gtrsim 0.45\,\varepsilon/k_B\approx 1.3\,T_g$.

\section{Conclusions}

In summary, the effect of thermal loading on mechanical and
structural properties of binary glasses was examined using molecular
dynamics simulations. The amorphous material was represented by the
binary Lennard-Jones mixture, which was initially annealed with a
computationally slow cooling rate from the liquid state to a
temperature well below the glass transition point.  It was shown
that a single cycle of heating and cooling leads to either
rejuvenated or relaxed states depending on the maximum temperature
and loading period. More specifically, the higher energy states were
obtained when the maximum temperature is above the glass transition
and the loading period is relatively small, so that the effective
heating/cooling rates are higher than the initial cooling rate.  In
contrast, thermal loading during longer periods resulted in lower
energy states due to either aging below $T_g$ or slower cooling from
the liquid state. The structural changes due to thermal loading are
reflected in the height of the first two peaks in the radial
distribution function of small atoms.  Moreover, the dependence of
the elastic modulus and the stress overshoot on the maximum loading
temperature is inversely related to the change in the potential
energy for a given loading period.  Thus, the simulation results
indicate that a significant decrease in the yield stress is obtained
in binary glasses rapidly heated and cooled above the glass
transition temperature at constant pressure conditions.

\section*{Acknowledgments}

Financial support from the National Science Foundation (CNS-1531923)
is gratefully acknowledged. The article was prepared within the
framework of the HSE University Basic Research Program and funded by
the Russian Academic Excellence Project `5-100'. The molecular
dynamics simulations were performed using the LAMMPS software
developed at Sandia National Laboratories~\cite{Lammps}. The
numerical simulations were performed at Wright State University's
Computing Facility and the Ohio Supercomputer Center.

%%%%%%%%%%%%%%% FIGURES %%%%%%%%%%%%%%%%%%%%%%%

% snapshot of the system
%
\begin{figure}[t]
\includegraphics[width=10.0cm,angle=0]{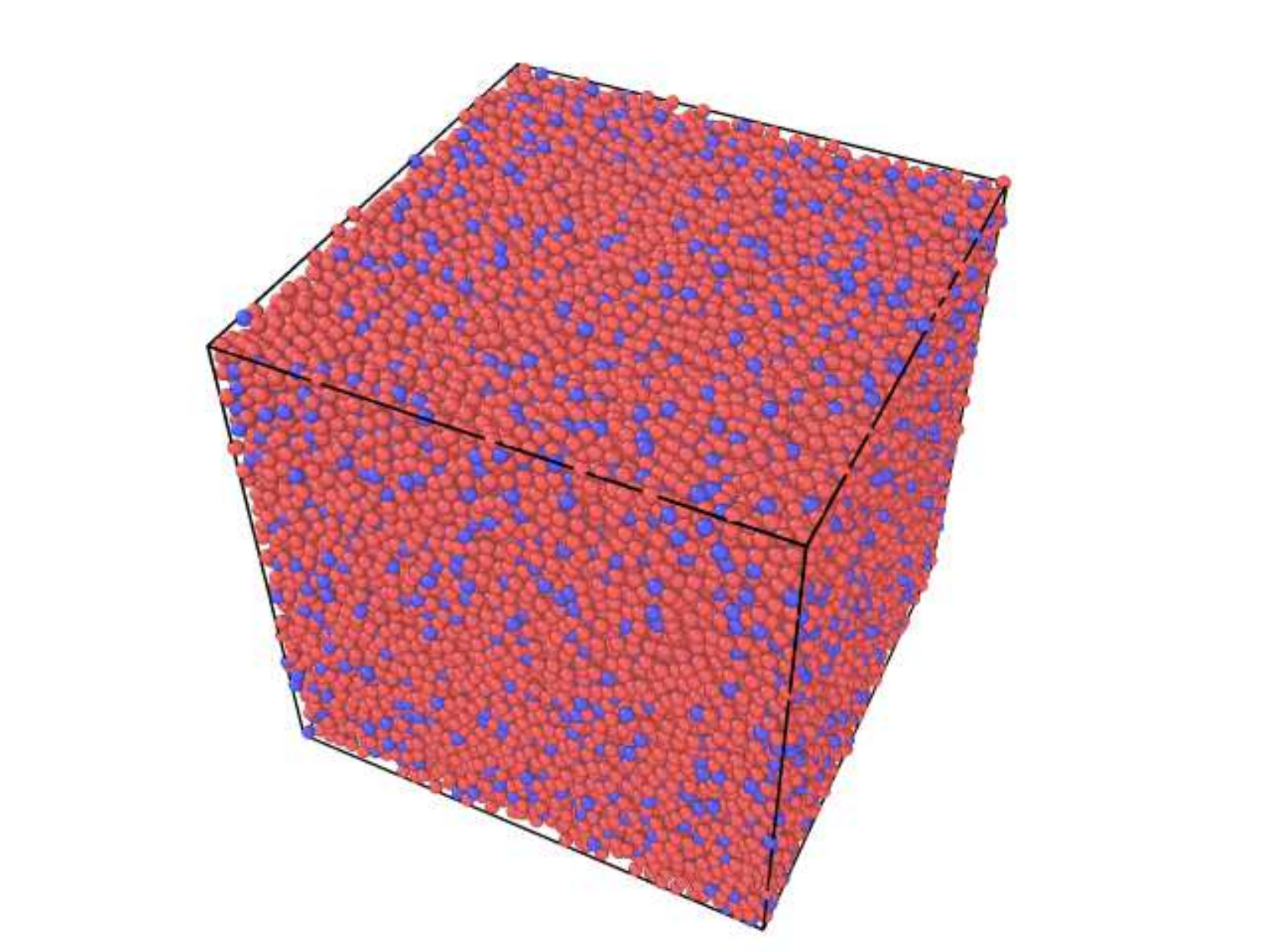}
\caption{(Color online) The configuration of atoms ($N=60\,000$) in
the binary glass after slow annealing at zero pressure to the
temperature $T_{LJ}=0.01\,\varepsilon/k_B$ with the cooling rate of
$10^{-5}\varepsilon/k_{B}\tau$. The atoms of types $A$ and $B$ are
denotes by red and blue spheres. The atoms are not depicted to
scale. The periodic boundary conditions are applied in three
dimensions. }
\label{fig:snapshot_system}
\end{figure}

% temperature control
%

% potential energy for sample with cooling rate 10^-5  T=5000\,\tau
%
%

% potential energy for sample with cooling rate 10^-5  T=10000\,\tau
%
%

% potential energy for sample with cooling rate 10^-5  T=500,000\,\tau
%
%

% U_1 versus TLJ
%

% radial distribution function gr_BB
%

% stress-strain curves
%

% yield stress and elastic modulus vs maximum temperature
%

\bibliographystyle{prsty}

\end{document}